\documentclass[10pt,letterpaper]{article}
\usepackage[top=0.85in,left=2.75in,footskip=0.75in]{geometry}

\usepackage{amsmath,amssymb}

\usepackage{colortbl}
\usepackage{float}
\usepackage{supertabular}
\usepackage{rotating}
\usepackage{booktabs}
\usepackage{ragged2e}
\usepackage[longtable]{multirow}
\usepackage{longtable}
\usepackage{array}
\usepackage{graphicx}
\usepackage{verbatim} 
\usepackage{interval}
\usepackage{amsmath,amssymb,amsfonts}
\usepackage{algorithmic}
\usepackage{subfigure}
\usepackage{adjustbox}
\usepackage{caption}
\usepackage{textcomp}
\usepackage{multicol}
\usepackage{setspace}
\restylefloat{figure}
\usepackage[hidelinks]{hyperref}

\usepackage{epstopdf}
\AppendGraphicsExtensions{.tif}

\usepackage{changepage}

\usepackage{textcomp,marvosym}

\usepackage{cite}

\usepackage{nameref,hyperref}

\usepackage[right]{lineno}

\usepackage[nopatch=eqnum]{microtype}
\DisableLigatures[f]{encoding = *, family = * }

\usepackage[table]{xcolor}

\usepackage{array}

\newcolumntype{+}{!{\vrule width 2pt}}

\newlength\savedwidth

\raggedright
\usepackage[aboveskip=1pt,labelfont=bf,labelsep=period,justification=raggedright,singlelinecheck=off]{caption}

\makeatletter
\renewcommand{\@biblabel}[1]{\quad#1.}
\makeatother

\usepackage{lastpage,fancyhdr,graphicx}
\usepackage{epstopdf}
\fancyheadoffset[L]{2.25in}
\fancyfootoffset[L]{2.25in}
\begin{document}
\vspace*{0.2in}

\begin{flushleft}
{\Large
\textbf\newline{Exploring Internet of Things Adoption Challenges in Manufacturing Firms: A Delphi Fuzzy Analytical Hierarchy Process Approach} 
}
\newline
\\
Hasan Shahriar\textsuperscript{1\ddag},
Md. Saiful Islam\textsuperscript{1},
Md Abrar Jahin\textsuperscript{1\ddag},
Istiyaque Ahmed Ridoy\textsuperscript{2},
Raihan Rafi Prottoy\textsuperscript{1},
Adiba Abid\textsuperscript{1},
and M. F. Mridha\textsuperscript{3*}

\bigskip
\textbf{1} Department of Industrial Engineering and Management, Khulna University of Engineering and Technology (KUET), Khulna 9203, Bangladesh
\\
\textbf{2} Institute of Business Administration, University of Dhaka, Dhaka, Bangladesh
\\
\textbf{3} Department of Computer Science, American International University-Bangladesh (AIUB), Dhaka 1229, Bangladesh
\\
\bigskip

%
%
\ddag Co-first author(s): Hasan Shahriar and Md Abrar Jahin. These authors contributed equally to this work.


* firoz.mridha@aiub.edu

\end{flushleft}


\begin{justify}
\section*{Abstract}
Innovation is key to gaining a sustainable edge in an increasingly competitive global manufacturing landscape. For Bangladesh's manufacturing sector to survive and thrive in today's cutthroat business environment, adopting transformative technologies such as the Internet of Things (IoT) is not a luxury but a necessity. This article tackles the formidable task of identifying and comprehensively evaluating the impediments to IoT adoption in the Bangladeshi manufacturing industry. We delve deeply into the complex terrain of IoT adoption challenges by synthesizing expert insights and a meticulously selected body of contemporary literature. We employ a robust methodology combining the Delphi method with the fuzzy Analytical Hierarchy Process to systematically analyze and prioritize these challenges. Using this methodology, we leveraged the combined expertise of domain specialists and subsequently employed fuzzy logic techniques to address the inherent ambiguities and uncertainties within the data. Our findings highlight this clear path. They reveal that among the myriad barriers, ``Lack of top management commitment to implementing new technology" (B10), ``High initial implementation investment costs" (B9), and ``Risks associated with switching to a new business model" (B7) loom most extensive, demanding immediate attention. These insights are not confined to academia but serve as a pragmatic guide for industrial managers. Armed with the knowledge gleaned from this study, managers can craft tailored strategies, set well-informed priorities, and embark on a transformational journey toward harnessing the vast potential of IoT in the Bangladeshi industrial sector. This article provides a comprehensive understanding of IoT adoption challenges and industry leaders with the tools necessary to navigate these challenges effectively. This strategic navigation, in turn, contributes significantly to enhancing the competitiveness and sustainability of Bangladeshi manufacturing in the IoT era.



\section*{Introduction}
\label{sec:introduction}
In the modern world, along with essential needs such as food, water, clothing, and shelter, the Internet has emerged as a transformative force, revolutionizing various industries. The Internet of Things (IoT) is a revolutionary concept that has gained substantial traction since its inception in 1999 when Kevin Ashton coined the term \cite{ashton_that_2009}. The IoT refers to a fast-expanding network of interconnected physical objects, including machines, automobiles, home appliances, and other items, enabling the exchange of data through connectivity, electronics, software, and sensors. With its potential to reshape how we live, work, and interact with the world, IoT stands out as a game-changer in today's technologically advanced era.

The impact of the IoT is already evident across multiple industries, from manufacturing and healthcare to transportation and agriculture. The IoT plays a significant role in manufacturing, enhancing operations, and driving competitiveness. Through IoT-enabled devices and real-time data exchange, manufacturing companies can implement predictive maintenance strategies, identify potential issues in advance, and optimize equipment performance, reducing downtime, increasing reliability, and prolonging equipment lifespan.

Moreover, IoT empowers manufacturers to streamline their supply chains (SC), tracking the movement of components, goods, and raw materials to improve inventory management and reduce wastage. They also foster energy efficiency by monitoring energy consumption and automating operations, leading to more sustainable and eco-friendly manufacturing practices. In addition, IoT enhances workplace safety by continuously monitoring environmental factors and alerting workers to potential hazards. The IoT drives process automation, enabling the industrial sector to lower labor costs and boost productivity. With the limitless possibilities of IoT, from simple temperature sensors to complex smart systems controlling entire factories, it has the potential to revolutionize the manufacturing industry by making companies more efficient, competitive, and responsive to market demands. However, as the number of connected devices increases, so does the importance of ensuring security and safeguarding the data generated. Cyber threats pose a significant concern, requiring manufacturers to prioritize robust security measures to protect sensitive information.


The IoT plays a significant role in the manufacturing industry, helping companies improve their operations and increase their competitiveness. There are several ways in which IoT is being used in the manufacturing industry:\\
\textit{Predictive maintenance}: The performance and health can be tracked with IoT devices, which also provide real-time data that can be utilized to identify possible issues before they arise. This can reduce downtime, increase equipment dependability, and increase the lifespan of the equipment.\\
\textit{Quality Control}: Manufacturing equipment performance and health can be tracked with IoT devices, which also provide real-time data that can be utilized to identify possible issues before they arise. This can reduce downtime, increase equipment dependability, and increase the lifespan of the equipment.\\
\textit{SC Optimization}: IoT can track the movement of components, completed goods, and raw materials along the SC, helping businesses manage their inventories more effectively and reduce waste.\\
\textit{Energy Efficiency}: By tracking energy use, finding areas for improvement, and automating operations to reduce energy use, IoT can assist manufacturing organizations in increasing their energy efficiency.\\
\textit{Safety}: IoT can increase workplace safety by monitoring factors such as temperature and humidity and warning workers of potential hazards.\\
\textit{Process Automation}: IoT automation streamlines manufacturing processes, lowers labor costs and increases productivity.

However, IoT in manufacturing presents challenges. Manufacturers need to understand the benefits of IoT and integrate it effectively into their operations. Compatibility issues with existing systems and associated expenses can hinder adoption, especially for smaller organizations. Securing IoT devices and data against cyber threats is crucial.

A notable research gap emerged across the reviewed studies, particularly in the manufacturing sector. While some studies have addressed aspects of manufacturing, such as IoT adoption barriers and evaluation models for industrial energy conservation, there remains a lack of comprehensive understanding in these areas. For instance, the emphasis on IoT barriers neglects crucial insights into enablers and opportunities within the broader scope of manufacturing SC IoT adoption. Similarly, the evaluation of Manufacturing Services (MS) credit lacks consideration for vital factors such as manufacturer distance, factory layout, and the introduction of new services, which are essential components in understanding the manufacturing network. Therefore, there is a clear need for more inclusive research that addresses both challenges and opportunities, specifically within the manufacturing sector, to ensure a holistic exploration of the impact of emerging technologies on industrial processes and SCs.

To address these challenges, this study aims to identify barriers to IoT adoption in the manufacturing sector using the Delphi method, evaluate and prioritize them using the Fuzzy Analytical Hierarchy Process (FAHP), and provide practical recommendations to overcome these obstacles and foster successful industry integration.

This study introduces novel contributions to the field by:

\begin{enumerate}
    \item An integrated framework that combines the Delphi method and FAHP provides a novel methodological approach to comprehensively assess IoT adoption barriers.
    
    \item Identifying barriers to IoT adoption in Bangladeshi manufacturing through the Delphi method.

    \item Systematically evaluating and prioritizing these barriers using the Fuzzy Analytical Hierarchy Process.

    \item Developing a dynamic prioritization model that adapts to the evolving nature of IoT adoption challenges, ensuring the relevance of prioritized barriers in the rapidly changing technological and business environments.
    
    \item Offering strategic recommendations tailored to the local context of Bangladeshi manufacturing addresses not only generic challenges but also specific socio-economic and industrial factors that influence IoT implementation.

    \item Validating the proposed framework through an empirical case study within the Bangladeshi manufacturing sector and demonstrating its practical applicability and effectiveness in addressing real-world challenges.
    
\end{enumerate}

The novelty of this study lies in its comprehensive approach, which combines the Delphi method and FAHP in the Bangladeshi manufacturing industry. While the Delphi method harnesses the collective wisdom of experts to identify and select the most relevant barriers to IoT adoption, FAHP considers these selected barriers and rigorously assesses their relative importance, accounting for the inherent uncertainty in decision-making. This dual-method approach provides a robust and nuanced understanding of the obstacles to IoT adoption in the Bangladeshi manufacturing sector, thereby enhancing the accuracy and practicality of the findings. These outcomes will empower companies to make informed decisions, develop effective implementation plans, and ultimately thrive in the increasingly competitive global business landscape driven by IoT innovations.

In the subsequent sections, we review the existing literature for the identification of barriers in the manufacturing industries in the ``\hyperref[sec:litrev]{Literature review}" section and then delve into the specifics of our research methodology, starting with the ``\hyperref[methods]{Methodology}" section. Here, we outline our approach, which includes three key subsections: ``Identifying Barriers for IoT Adoption," ``Fuzzy Delphi Method" for selecting relevant barriers, and the ``FAHP Methodology" for ranking these barriers. Following this, we present our findings and engage in comprehensive discussions in the ``\hyperref[results]{'Results and discussions'}" section, in which we provide actionable recommendations tailored to manufacturing firms aiming to integrate IoT technologies into their processes, emphasizing the importance of top management commitment and strategic planning. In the ``\hyperref[sec:real-world]{Real-world applications and practical implications}" section, we showcased authentic case studies of companies such as Siemens AG and General Electric, demonstrating how IoT technologies have been successfully implemented to address operational challenges and drive efficiency improvements in the manufacturing sector. We then draw key insights and implications in the ``\hyperref[conclusion]{'Conclusions and future works'}" section, shedding light on the critical barriers to IoT adoption in Bangladesh's manufacturing sector. Finally, we outline avenues for future research by providing a holistic view of the structure and objectives of our study.

\section*{Literature review}
\label{sec:litrev}
The inception of IoT can be traced back to the early 1990s when the idea of interconnected devices began to take shape at MIT's Auto-ID Center. Kevin Ashton, the Center's director, introduced the term ``IoT" in 1999~\cite{greengard_internet_2015}. The concept revolved around using radio-frequency identification (RFID) tags to track products along the SC of companies such as Procter and Gamble. As early as 1997, Ashton envisioned the potential of RFID tags to scan and identify items with data transmitted wirelessly across a network. Notably, the industry began implementing RFID tags as early as 1980 \cite{xu_internet_2014}. Subsequently, the idea of a wireless sensor network (WSN) emerged, finding applications in traffic management and healthcare. This new concept incorporates sensors and actuators to detect, track, and monitor various objects \cite{xu_internet_2014}. Over time, this vision has evolved to encompass other technologies, such as GPS devices, cell phones, social networks, cloud computing, and data analytics, forming the foundation for modern IoT concepts. IoT has emerged as a pivotal technology driving the development of Industry 4.0, particularly in the European manufacturing industry, with Germany leading the charge. Industry 4.0 represents the fourth industrial revolution, building upon its predecessors, Industry 1.0, Industry 2.0, and the Digital Revolution (Industry 3.0). According to Zhou et al.\cite{zhou_industry_2015}, Industry 4.0 integrates information and communications technologies seamlessly with industrial processes. To realize the vision of Industry 4.0, IoT serves as a critical enabler alongside cloud manufacturing and cyber-physical systems (CPS). Cyber-physical systems consist of intelligent equipment, storage systems, and production facilities capable of autonomous communication, action, and monitoring \cite{kagermann_secure_2012}. CPS involves the integration of analog/digital hardware, bridging manufacturing entities' physical and virtual components \cite{cheng_industry_2016}. The IoT provides the necessary framework to connect CPS via a network of devices, sensors, and actuators. In this context, cloud manufacturing (CM) has emerged, leveraging cloud computing resources hosted in external data centers to complement the initiatives of Industry 4.0. Together, these technological advancements promise to revolutionize manufacturing, enabling enhanced connectivity, data-driven decision-making, and increased efficiency in industrial processes. As the IoT continues to advance and intertwine with other technologies, it is expected to play an even more significant role in reshaping the manufacturing landscape, unlocking further possibilities for innovation and growth.

In their article, Cui et al.\cite{cui_internet_2021} devised a framework for evaluating IoT adoption barriers in a circular economy. They integrated Step-wise Weight Assessment Ratio Analysis (SWARA) and Combined Compromise Solution (CoCoSo) methods based on Pythagorean Fuzzy Sets. SWARA assessed the significance of barriers, while CoCoSo ranked organizations under these barriers. These methods demonstrated realistic outcomes, thereby emphasizing its applicability. The identified barriers underscored the need to explore other technologies, such as RFID, cloud computing, big data analytics, and cybersecurity in the circular economy. Rizwan et al.\cite{rizwan_internet_2022} enhanced SC network performance with edge computing and prioritized data access using Intelligent K-means clustering. Low-priority data were logged, and the high-priority access smart contract ensured a secure blockchain data flow. The Python-based approach showed improved scalability, reaction time, throughput, and accuracy, outperforming the existing blockchain technology in SC management. However, integrating blockchain in IoT faced limitations, prompting consideration of permissioned blockchains and ``Blockchain pruning" for scalability and privacy concerns.

Kamble et al.\cite{kamble_analysis_2018} utilized Interpretive Structural Modeling (ISM) to establish relationships among barriers to Industry 4.0 adoption. Fuzzy Cross-impact matrix multiplication applied to classification (MICMAC) analysis determined the driving and dependence power, incorporating input from industry and academia experts. The findings aid in classifying significant barriers and understanding their effects on adoption. Practical insights can support practitioners and policymakers in building a valid digital manufacturing platform. Sumrit\cite{sumrit_evaluating_2022} designed an Industrial Internet of Things (IIoT) adoption evaluation framework with Interval-Valued Pythagorean Fuzzy Sets to address decision-making uncertainty. Twelve validated readiness criteria, determined by IVPF-AHP, inform manufacturers about their readiness and guide early actions for successful adoption; however, the study's limitation is its exclusive focus on barriers, neglecting enablers and opportunities in manufacturing SC IoT adoption.

Shijie \& Yingfeng\cite{shijie_credit-based_2021} utilized the fuzzy Analytic Network Process and Cross-Entropy to evaluate Manufacturing Services (MS) credit in complex manufacturing networks. The proposed Service Scoring Mechanism (SSM) enables real-time task allocation based on credit, enhancing decision-making and customer satisfaction. However, this study primarily focused on the evaluation of MS. It does not consider factors such as the manufacturer distance, factory location layout, nodal elimination, and introduction of new services in the manufacturing network. Chen\cite{chen_intelligent_2020} proposed a smart factory architecture using industrial IoT technology for manufacturing workshops, addressing challenges like diverse data and strong correlations. The proposed system, incorporating WSN and RFID, demonstrated effective real-time data collection and product tracking. However, the study emphasized that achieving intelligent manufacturing requires further exploration of manufacturing IoT technology, artificial intelligence algorithms, and machine learning beyond the initial real-time monitoring. Mehbodniya\cite{mehbodniya_energy-aware_2022} proposed a Multilayer Energy-Aware RPL (MCEA-RPL) cluster to optimize IoT networks, reduce data traffic, and extend lifespan in three phases. Blockchain technology enhances the network's lifetime by minimizing identical data transfers. Enhanced Mobility Support RPL (EM-RPL) in Industrial IoT improves mobility support with blockchain, thereby efficiently reducing route interruptions. This study employs a limited number of sink nodes for network information collection. Alowaidi\cite{alowaidi_fuzzy_2022} developed a fuzzy architecture to optimize intelligent microgrids with storage for sustainability and regulated loads, utilizing solar power, wind speed, and power load parameters. It enhances electricity and storage efficiency, providing insights into microgrid management through an expert-based system. However, this study focused solely on residential renewable energy management, omitting integration challenges with larger grid infrastructure.

Kumar et al.\cite{kumar_iiot_2021} identified ten IIoT implementation challenges from the literature and expert input. Researchers surveyed Indian manufacturing SMEs using a questionnaire and categorized the challenges using the DEMATEL technique. They proposed using blockchain to address challenges such as data security and technology reliability, highlighting the potential benefits for Indian SMEs despite the limited sample size and geographical focus. Khanna \& Kaur\cite{khanna_internet_2020} undertook a comprehensive review of the literature on different aspects of IoT, encompassing technologies, applications, and challenges, to evaluate researchers' contributions and highlight existing gaps. Future research directions are suggested to guide newcomers and advance the field with innovative ideas. Kalsoom et al.\cite{kalsoom_impact_2021} conducted a detailed review to explore IoT applications in manufacturing Industry 4.0 to fill gaps. They used a systematic literature review method, identified key gaps, found six differences between traditional and Industry 4.0, and outlined ten enablers and 11 challenges. Finally, 11 research areas were proposed to address these gaps. He et al.\cite{he_development_2021} developed a scale to evaluate the competencies of clinical nursing teachers using the Delphi method and then tested the reliability and validity of the resulting assessment scale. They used a convenience sampling method to select participants for testing the scale, and this sampling approach slightly influenced the factor selection process during scale development. Ehie \& Chilton\cite{ehie_understanding_2020} highlighted the crucial role of IoT in digital manufacturing, noting that over two-thirds of manufacturing executives anticipate its potential to enhance competitive advantage. However, many companies lack a comprehensive understanding of IoT adoption. To facilitate effective IoT adoption, they developed and tested a two-stage model grounded in information technology/operational technology convergence. This model offers a practical roadmap for manufacturing organizations adopting IoT technologies. Ullah et al.\cite{ullah_twenty-one_2020} aimed to create an unbiased method for organizations to choose the appropriate IoT platform tailored to their needs. This was achieved by examining IoT fundamentals, identifying key platform factors, and testing a selection framework with five examples. However, the study was limited to theoretical framework testing, and future research is needed to assess platforms across different industries and to develop an automated process for factor weighting.

Li et al.\cite{li_fuzzy_2017} constructed a comprehensive evaluation model to monitor industrial energy conservation. An evaluation index for industrial energy-intensive equipment was proposed. Subsequently, by integrating an analytic hierarchy process and a fuzzy comprehensive evaluation method, an IoT-based industrial EMS can evaluate the operational level of energy-intensive equipment fully. Enhancing IoT-based EMS faces security, scalability, and accuracy challenges, necessitating advanced hardware, algorithms, and a tailored data collection scheme for specialized energy conservation projects within the integrated context of cloud computing and big data. Yuen et al.\cite{yuen_intelligent_2019} introduced an EM-RMM to assess IoT-related risks in manufacturing plants, utilizing FAHP to analyze risk likelihood and consequences. A case study on environmentally sensitive electronics manufacturing systematically assessed IoT implementation risks for achieving smart manufacturing. However, the model relied on expert assessments, lacked objective data-driven analysis, and required broader validation across diverse manufacturing contexts. Kumar et al.\cite{kumar_narrowing_2020} pinpointed barriers to adopting Industry 4.0, highlighting the lack of transparent cost-benefit analysis as the main obstacle. Overcoming these barriers, especially addressing cost-benefit concerns, is crucial for motivating businesses. These insights guide managers and policymakers; however, real-world testing is needed for further research. Khan et al.~\cite{khan_unearthing_2023} examined barriers to IoT adoption in the Food SC (FSC), identifying 14 through a literature review. Twelve significant barriers were determined using the fuzzy Delphi method with expert input, including complex frameworks, IoT affordability, and poor IT infrastructure. The findings offer insights for FSC managers to digitize their SC by addressing past barriers. However, the research is limited to IoT deployment in FSC and excludes other SC types.

IoT implementation in the manufacturing sector presents several challenges that hinder seamless integration and utilization. These obstacles must be carefully addressed in order to harness the advantages of the IoT entirely in industrial settings. The following are the key barriers manufacturers may encounter:\\
\textit{Technical complexity}: The implementation of IoT technology in manufacturing requires high technical expertise and understanding. The complexity of IoT systems and their interfaces can be daunting for manufacturers, making it challenging for them to grasp the full potential and benefits of IoT. To overcome this hurdle, educational initiatives and training programs can be employed to empower manufacturers with the necessary technical knowledge to effectively embrace and leverage IoT.\\
\textit{Integration Challenges}: Integrating IoT technology into existing manufacturing systems can be arduous, particularly if these systems are outdated or not designed with the IoT in mind. The lack of compatibility may lead to issues with seamless communication and operation between different components, impeding the overall efficiency and performance of the IoT implementation. Manufacturers should thoroughly assess their current systems and strategize for a smooth transition involving gradual upgrades or replacements to ensure harmonious integration.\\
\textit{Cost Considerations}: The adoption of IoT technology often results in significant expenses, which may pose a particular challenge for small and medium-sized organizations with limited financial resources. The costs associated with the IoT include investments in hardware, software, installation, and ongoing maintenance. To mitigate this barrier, manufacturers can explore cost-effective IoT solutions, consider partnerships, or seek government incentives and support to facilitate the adoption of IoT.\\
\textit{Data Security and Privacy}: As IoT devices generate and transfer vast amounts of sensitive data during manufacturing, ensuring the security and privacy of the information becomes critical. The potential for cyberattacks and unauthorized access necessitates robust security measures, encryption protocols, and data management practices. Manufacturers must prioritize data protection and establish comprehensive cybersecurity strategies to safeguard their IoT systems and the valuable data that they collect.\\
\textit{Lack of Universal Standardization}: There is a universal standard for IoT systems and devices, leading to a fragmented market with numerous solutions and technologies. This lack of standardization can make it challenging for manufacturers to select the most suitable IoT solutions for their specific needs and to ensure seamless interoperability between various devices and platforms. Industry can address this barrier by fostering collaboration and developing common standards to facilitate a more cohesive and integrated IoT ecosystem.\\
\textit{Reluctance to Adopt New Processes}: Some manufacturers may hesitate to adapt their established procedures and infrastructure to accommodate IoT technologies. This reluctance can impede the widespread adoption of IoT, preventing companies from fully realizing their advantages. Overcoming this barrier requires effective change management strategies, clear communication of the benefits, and successful case studies to demonstrate the positive impact of IoT implementation.

The successful integration of the IoT in the manufacturing sector depends on overcoming these barriers. By addressing technical complexities, facilitating seamless integration, managing costs, prioritizing data security, fostering standardization, and promoting a positive shift in mindset, manufacturers can embrace IoT's potential and achieve enhanced efficiency, innovation, and competitiveness in their operations.

\section*{Methodology}
\label{methods}
Our study employed a three-phase methodology to ascertain and rank the primary factors that inhibit IoT adoption in manufacturing firms. The initial phase entailed an extensive review of the existing literature, and the following phase entailed expert insights employing the Delphi method to pinpoint the crucial criteria. In the final phase, we leveraged the widely utilized FAHP to evaluate and rank these criteria in terms of their significance. This distinctive hybrid approach, depicted in Fig \ref{fig1}, explains a comprehensive methodological flowchart for identifying and ranking the influential barriers to  IoT adoption in manufacturing firms.

\begin{figure*}[!ht]
\centering
\includegraphics[width=1\linewidth, height=1\linewidth]{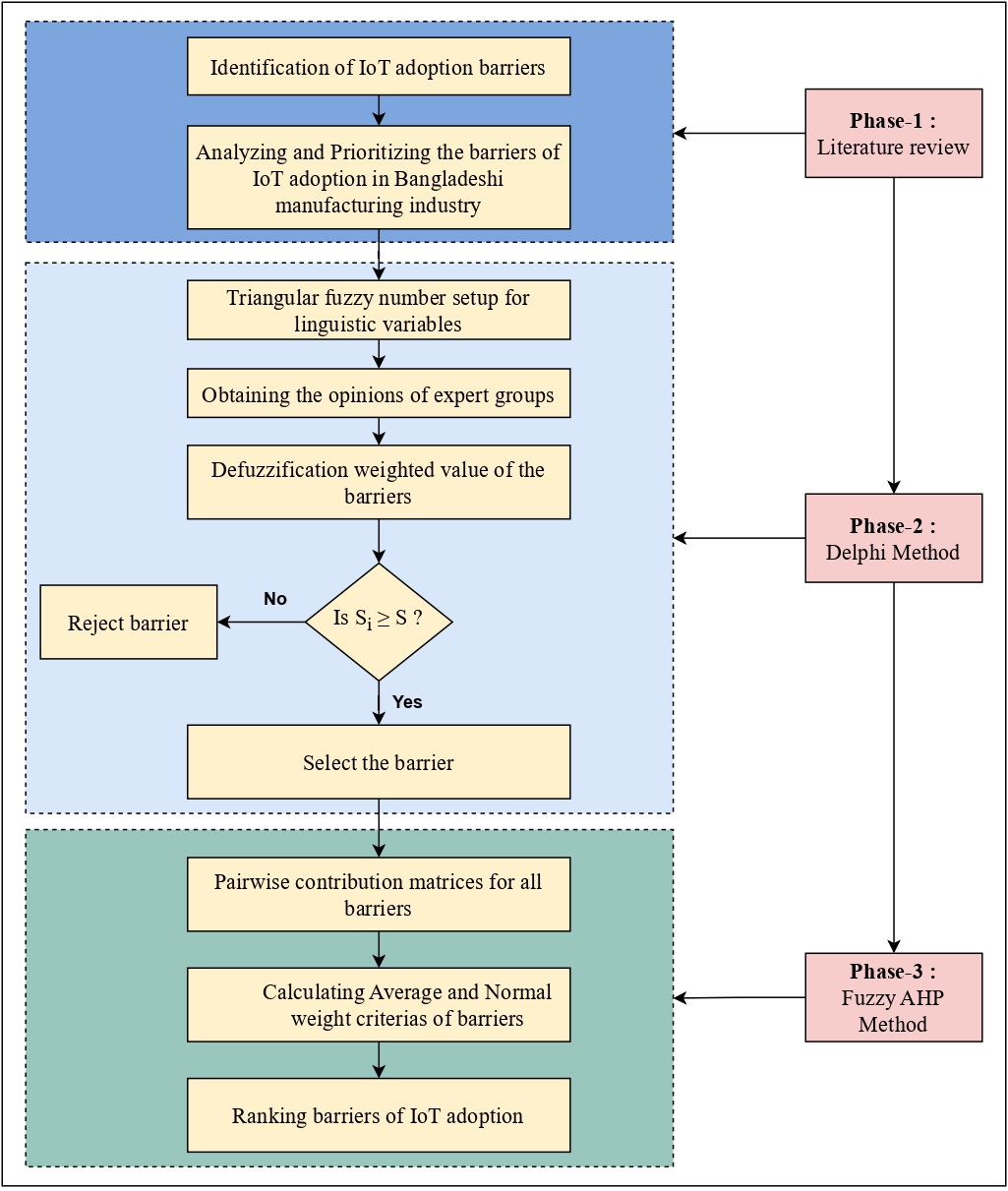}
\caption{\bf Flowchart of our proposed research approach.}
\label{fig1}
\end{figure*}

\subsection*{Identifying barriers for IoT adoption}
This study identifies barriers to IoT adoption in the manufacturing sector based on a comprehensive review of relevant research papers. The search process employed three significant keywords: (``Internet of Things" OR ``IoT") AND ``adoption" AND ``manufacturing" AND (``industry" OR ``sector)" to ensure a targeted and focused approach. A rigorous search was conducted using the Google Scholar database to gather research articles, encompassing a diverse range of peer-reviewed journals. It is important to note that white papers, theses, book chapters, conference proceedings, and websites were excluded from this evaluation, including only high-quality, peer-reviewed literature. Additionally, for consistency and accessibility, only articles published in English were included in this study. This decision ensured that the language barrier did not affect the comprehensive analysis and interpretation of the chosen articles. Furthermore, a meticulous Title and Abstract review process was undertaken to select articles that directly pertained to the objectives of this study. Articles that were not directly aligned with the research focus were excluded from the final set of selected papers. Moreover, additional papers were included after thoroughly examining the reference lists of the selected articles to enhance the comprehensiveness of the identified barriers. This approach helped capture potential barriers that might have been explored in related studies but not explicitly mentioned in the initial search results.

The culmination of this systematic process resulted in the identification of 16 significant barriers to IoT adoption in the manufacturing sector. As outlined in the research objectives, these barriers will serve as the foundation for subsequent evaluation and prioritization using the FAHP). Table \ref{tab1} presents a concise overview of the identified barriers and briefly describes each, setting the stage for further analysis and discussion.


\begin{sidewaystable*}
\centering
\caption{\bf IoT adoption barriers in Bangladeshi manufacturing industry.}
\resizebox{\linewidth}{!}{%
\footnotesize
\begin{tabular}{|>{\centering\hspace{0pt}}m{0.04\linewidth}|>{\hspace{0pt}}m{0.137\linewidth}|>{\hspace{0pt}}m{0.413\linewidth}|>{\hspace{0pt}}m{0.348\linewidth}|} 
\hline
\textbf{Serial No} & \textbf{Barriers for IoT Adoption} & \textbf{Description} & \textbf{Authors} \\ 
\hline
B1 & Absence of training initiatives & After implementing IoT, staff members will no
  longer be only operators; instead, they will be real problem solvers who
  analyze the data IoT has collected. Training programs will be required in order to grasp IoT technology. & \cite{kiel_influence_2017, ribeiro_da_silva_pursuit_2019, chen_intelligent_2020} \\ 
\hline
B2 & Lack of knowledge of return on investment
  (ROI) & Adoption of IoT technology is hampered by the
  lack of a standardized method for calculating ROI, particularly when a
  significant initial investment is required. Evaluating and producing
  financial returns involves high risk. The largest obstacle to IoT adoption is
  an unknown payback period. & \cite{chan_determinants_2013, lim_rfid_2013, chang_determinants_2008} \\
\hline
B3 & Interoperability issues & IoT devices have trouble communicating with
  one another since they are made by many vendors and adhere to various
  standards. The use of IoT devices may be constrained, and this lack of interoperability may slow adoption. & \cite{kiel_influence_2017, asir_internet_2015} \\ 
\hline
B4 & Unawareness of the advantages of IoT & Despite the fact that there are many 
  articles discussing the benefits of using IoT technology, many IoT
  applications are still in their infancy, and the benefits of the technology
  for businesses are not well understood. & \cite{fosso_wamba_determinants_2016, haddud_examining_2017, chan_determinants_2013, tu_exploratory_2018} \\ 
\hline
B5 & Employees' resistance to embracing new
  technologies & The fear of losing their jobs prevents
  workers from embracing IoT technologies. & \cite{silva_requirements_2019, haddud_examining_2017} \\ 
\hline
B6 & Limited connectivity & A dependable and steady internet connection
  is necessary for IoT devices to operate correctly. Adoption may be hampered
  in some places by low or sporadic connectivity. & \cite{silva_requirements_2019, chen_empirical_2012} \\ 
\hline
B7 & Power and battery life & Many IoT gadgets run on batteries and need to
  be charged frequently, which can be annoying for consumers. In some
  circumstances, it may be difficult to use particular devices because they
  need a steady power source. & \cite{chan_determinants_2013, lim_rfid_2013, chang_determinants_2008} \\ 
\hline
B8 & Regulatory challenges & IoT devices may be governed by several
  laws, which can differ by sector, nation, or location. Businesses may find it
  difficult to comply with these rules, which could prevent adoption. & \cite{silva_requirements_2019}, \cite{haddud_digitalizing_2020} \\ 
\hline
\end{tabular}
}
\label{tab1}
\end{sidewaystable*}

\begin{sidewaystable*}
\centering
\caption*{\bf Continuation of Table \ref{tab1}: IoT adoption barriers in Bangladeshi manufacturing industry.}
\resizebox{\linewidth}{!}{%
\footnotesize
\begin{tabular}{|>{\centering\hspace{0pt}}m{0.04\linewidth}|>{\hspace{0pt}}m{0.137\linewidth}|>{\hspace{0pt}}m{0.413\linewidth}|>{\hspace{0pt}}m{0.348\linewidth}|} 
\hline
\textbf{Serial No} & \textbf{Barriers for IoT Adoption} & \textbf{Description} & \textbf{Authors} \\ 
\hline
B9 & Security and privacy concerns & Data security is a serious worry because the
  IoT is constantly sharing massive amounts of data. To prevent
  risks to data security, authentication, authorization, and access control
  procedures are necessary. Data privacy is another key problem because these
  linked gadgets are exchanging a lot of personal or industry-specific
  information. & \cite{lin_integrated_2009}, \cite{lim_rfid_2013}, \cite{chan_determinants_2013}, \cite{asir_internet_2015}, \cite{hsu_understanding_2017}, \cite{reyes_determinants_2016}, \cite{kiel_influence_2017}, \cite{contreras_importance_2018}, \cite{haddud_digitalizing_2020}, \cite{silva_requirements_2019} \\ 
\hline
B10 & Unskilled or inexperienced Manpower & Handling IoT technologies requires skilled
  personnel. IoT technology installation requires technical expertise and IT
  skills due to the massive amount of generated data. & \cite{silva_requirements_2019}, \cite{haddud_digitalizing_2020} \\ 
\hline
B11 & Risks associated with switching to a new
  business model & Any technological paradigm shift necessitates
  modifying the business model. The implementation of IoT technology
  requires a new business model framework. Organizations are reluctant to
  assume the risk associated with implementing a new business model for IoT
  technology because they are unsure of the ROI. & \cite{haddud_examining_2017},\cite{kiel_influence_2017},\cite{arnold_how_2016} \\ 
\hline
B12 & Data overload & IoT devices can produce a lot of data, which
  can be daunting for both individuals and enterprises. This data can be
  difficult to process and analyze, especially for individuals lacking the
  knowledge or skills to do so. & \cite{wang_evolution_2021}, \cite{bi_internet_2014}, \cite{chan_determinants_2013}, \cite{asir_internet_2015} \\ 
\hline
B13 & Inadequate technology & Needs for IoT applications real-time analysis
  of data produced by IoT devices requires high-speed internet and IT gear to
  implement, support, and manage IoT devices. & \cite{bi_internet_2014}, \cite{pool_rfid_2015}, \cite{asir_internet_2015}, \cite{reyes_determinants_2016}, \cite{silva_requirements_2019}, \cite{lin_integrated_2009}, \cite{wang_evolution_2021}, \cite{bi_internet_2014} \\ 
\hline
B14 & High initial implementation investment costs & IoT implementation requires a significant
  upfront investment, which includes the cost of sensors, networks, software, maintenance,
  training, and other expenses. Indian organizations lack the necessary
  financial resources. Additionally, given that sensor costs are falling daily,
  Indian businesses prefer to wait for lower prices. & \cite{chang_determinants_2008, lin_integrated_2009, chen_empirical_2012, radziwon_smart_2014, chan_determinants_2013, asir_internet_2015, reyes_determinants_2016, haddud_examining_2017, tu_exploratory_2018, silva_requirements_2019} \\ 
\hline
B15 & Lack of top management's commitment to implementing new technology & The acceptance of new technology by a company is strongly influenced by top management backing. IoT adoption and implementation demands senior management support as well as time and financial
  commitment. & \cite{chang_determinants_2008},\cite{lin_integrated_2009},\cite{wang_evolution_2021},\cite{chan_determinants_2013},\cite{pool_rfid_2015},\cite{reyes_determinants_2016},\cite{hsu_understanding_2017},\cite{silva_requirements_2019} \\ 
\hline
B16 & Absence of Standardization & IoT implementation is difficult. Still, the greatest obstacle to this technology's acceptance is the lack of standardized interoperability interfaces between various IoT devices made by various suppliers. Various businesses have created versions of standards that are incompatible with one another. & \cite{chen_empirical_2012, chang_determinants_2008, lim_rfid_2013, bi_internet_2014, asir_internet_2015, haddud_examining_2017} \\
\hline
\end{tabular}
}

\end{sidewaystable*}

\subsection*{Fuzzy Delphi method to select the relevant barriers}
The Fuzzy Delphi Method is a technique for combining expert thoughts and opinions to reach a consensus on complex and individualized situations. It blends the conventional Delphi technique with fuzzy logic and sets a theory to deal with ambiguity and imprecision in expert opinions.

The Fuzzy Delphi approach entails several rounds of anonymous questionnaires and comments, where experts are invited to express their thoughts using language variables (e.g., very high, high, moderate, low, very low) rather than numerical values. Fuzzy set operations were used to aggregate expert opinions once these linguistic variables were converted into fuzzy sets using fuzzy membership algorithms.

\subsubsection*{The linguistic variables' triangular fuzzy number setup}
A fuzzy set with a triangular membership function is referred to as a triangular fuzzy set. The triangular fuzzy set formula is given by:

\begin{equation}
\begin{aligned}
    \mu_A(x) =
    \begin{cases}
        0, & \text{if } x < a \\
        \frac{x - a}{b - a}, & \text{if } a \leq x \leq b \\
        \frac{c - x}{c - b}, & \text{if } b \leq x \leq c \\
        0, & \text{if } x > c
    \end{cases}
    \end{aligned}
\end{equation}

A triangular fuzzy number $a_{ij}$ is defined as $a_{ij} = (a_{ij}, b_{ij}, c_{ij})$,  where $i \in \{1, 2, \ldots, n\}$ denotes the experts and $j \in \{1, 2, \ldots, m\}$ denotes the barriers, where $n$ and $m$ are the numbers of experts and barriers, respectively. The linguistic scales and their fuzzy equivalents are listed in Table \ref{tab2}.

\begin{table}
\centering
\caption{\bf Setting up fuzzy number set for linguistic variables.}
\begin{tabular}{|c|c|c|l} 
\cline{1-3}
\textbf{Linguistic Variables} & \textbf{Rating} & \textbf{Triangular Number} &   \\ 
\cline{1-3}
Very low                      & 1               & (0,0,1)                    &   \\ 
\cline{1-3}
Low                           & 2               & (1,2,3)                    &   \\ 
\cline{1-3}
Low medium                    & 3               & (2,3,4)                    &   \\ 
\cline{1-3}
Medium                        & 4               & (3,4,5)                    &   \\ 
\cline{1-3}
Medium High                   & 5               & (4,5,6)                    &   \\ 
\cline{1-3}
High                          & 6               & (5,6,7)                    &   \\ 
\cline{1-3}
High-very High                & 7               & (6,7,8)                    &   \\ 
\cline{1-3}
Very High                     & 8               & (7,8,9)                    &   \\ 
\cline{1-3}
Very High-Extreme             & 9               & (8,9,10)                   &   \\ 
\cline{1-3}
Extreme                       & 10              & (10,10,10)                 &   \\
\cline{1-3}
\end{tabular}
\label{tab2}
\end{table}

\subsubsection*{Obtaining the opinions of experts on the significance of the barriers}
At this stage, we gathered feedback from experts from relevant fields on the relevance and applicability of barriers. The expert consultation involved distributing questionnaires via Google Forms, and the participants' recruitment period was from 22 Jan 2023 to 17 Mar 2023. Confidentiality of participant responses was strictly maintained. All data collected were anonymized and stored securely. Only the research team has access to the raw data, and findings were reported in aggregate form to ensure the anonymity of participants. Participants were provided with informed consent forms detailing the purpose of the study, their rights as participants, and procedures for data handling. Participation in the survey was voluntary, and participants had the right to withdraw at any time without penalty. 

Table \ref{tab:demographic_profile} outlines the demographic composition of the selected experts, offering insights into their gender distribution, age profiles, experience levels, and areas of expertise. The data suggest a representation of 10 males and 6 females. In terms of age, a diverse range was observed, mainly falling within the 30 to 50-year-old range. Expertise varies, ranging from less than 5 years to over 15 years, indicating a blend of experience levels. Additionally, the experts bring diverse skill sets from 4 sectors, including Manufacturing, IT \& Cybersecurity, Industrial Systems Management, Academic IoT Research, and interdisciplinary fields. The Ethics Review Committee of the Research and Extension Center at Khulna University of Engineering \& Technology (KUET), Bangladesh, granted ethical approval, allowing us to proceed with our study involving expert participants.

A seven-point Likert scale was used to assess responses to all questions\cite{sullivan_analyzing_2013}. The choices were ``strongly agree," ``agree," ``more or less agree," ``undecided," ``more or less disagree," ``disagree," and ``strongly disagree". The project team convened the experts to personally clarify the research's objectives and importance, seeking their consent.

\begin{table}[htbp]
\centering
\caption{\bf Demographic profile of the selected experts}
\label{tab:demographic_profile}
\begin{tabular}{@{}ll@{}}
\toprule
\textbf{Demographics}                     & \textbf{No. of experts} \\ \midrule
\textbf{\textit{Gender}}                                     &                         \\
Male                                                         & 10                      \\
Female                                                     & 06                      \\ \midrule
\textbf{\textit{Age}}                                           &                         \\
Below 30 years                                         & 03                      \\
30 to 40 years                                           & 05                      \\
41 to 50 years                                           & 06                      \\
Above 50 years                                         & 02                      \\ \midrule
\textbf{\textit{Expertise in their fields}} &                         \\
Less than 5 years                                   & 02                      \\
5 to 10 years                                           & 06                      \\
10 to 15 years                                         & 06                      \\
Above 15 years                                       & 02                      \\ \midrule
\textbf{\textit{Domain of expertise}}         &                         \\
Manufacturing Professionals               & 02                      \\
IT \& Cybersecurity Specialists               & 03                      \\
Industrial Systems Management Experts & 04                      \\
Academic IoT Researchers                     & 06                      \\
Cross-Disciplinary Experts                      & 01                      \\ \bottomrule
\end{tabular}
\end{table}

Cronbach's alpha was used to check the consistency of the survey items. This helps researchers and practitioners ensure that they are measuring things accurately. Cronbach's alpha measures how well many measurements add up, such as scores on a test or questionnaire. We used a 240-item questionnaire to explore manufacturing firms' IoT adoption barriers, with each question rated on a scale from 0 to 6. A Cronbach's alpha value above 0.7 is considered good, and ours was 0.7226. The variance of the total score (3269.717) and the sum of the item variances (916.5667) provide this dependability measure, showing that our questionnaire is reliable. 

We conducted a Delphi study to elucidate the central barriers influencing the risk index associated with IoT adoption challenges in manufacturing firms. To ensure a comprehensive examination, we engaged the expertise of 16 distinguished professionals hailing from five distinct domains. We meticulously formed 4 four hybrid groups, each comprising four experts, all possessing advanced degrees and specialized knowledge in their respective fields. They were deliberately selected for their diverse perspectives and significant contributions to the discipline. Following a structured process involving an extensive literature review, collaborative brainstorming sessions, and rigorous application of the Delphi method, the experts within each group iteratively evaluated and prioritized the identified barriers. Through multiple rounds of surveys and deliberations, a consensus emerged within each group, identifying and assessing 16 barriers spanning the technological, organizational, and operational dimensions of IoT adoption in manufacturing firms. Thus, 24 expert opinions were synthesized from the collective insights of the four expert groups, as shown in Table \ref{tab:expert_group}.

\begin{table*}[!ht]
\begin{adjustwidth}{-2.25in}{0in}
\centering
\caption{\bf Gathered opinions from the expert groups on identifying relevant barriers.}
\resizebox{\linewidth}{!}{%
\begin{tabular}{|c|c|c|c|c|} 
\hline
\textbf{Barriers} & \textbf{Group-1 Expert Opinion} & \textbf{Group-2 Expert Opinion} & \textbf{Group-3 Expert Opinion} & \textbf{Group-4 Expert Opinion}  \\ 
\hline
B1                & (6,7,8)                   & (6,7,8)                   & (6,7,8)                   & (7,8,9)                    \\ 
\hline
B2                & (7,8,9)                   & (6,7,8)                   & (8,9,10)                  & (8,9,10)                   \\ 
\hline
B3                & (6,7,8)                   & (5,6,7)                   & (4,5,6)                   & (5,6,7)                    \\ 
\hline
B4                & (7,8,9)                   & (7,8,9)                   & (7,8,9)                   & (8,9,10)                   \\ 
\hline
B5                & (5,6,7)                   & (10,10,10)                & (7,8,9)                   & (7,8,9)                    \\ 
\hline
B6                & (5,6,7)                   & (4,5,6)                   & (5,6,7)                   & (3,4,5)                    \\ 
\hline
B7                & (6,7,8)                   & (4,5,6)                   & 2,3,4                     & (3,4,5)                    \\ 
\hline
B8                & (4,5,6)                   & (2,3,4)                   & (2,3,4)                   & (3,4,5)                    \\ 
\hline
B9                & (10,10,10)                & (8,9,10)                  & (8,9,10)                  & (10,10,10)                 \\ 
\hline
B10               & (8,9,10)                  & (6,7,8)                   & (7,8,9)                   & (6,7,8)                    \\ 
\hline
B11               & (6,7,8)                   & (9,10,10)                 & (7,8,9)                   & (7,8,9)                    \\ 
\hline
B12               & (1,2,3)                   & (2,3,4)                   & (5,6,7)                   & (4,5,6)                    \\ 
\hline
B13               & (10,10,10)                & (10,10,10)                & (7,8,9)                   & (6,7,8)                    \\ 
\hline
B14               & (8,9,10)                  & (8,9,10)                  & (6,7,8)                   & (7,8,9)                    \\ 
\hline
B15               & (10,10,10)                & (8,9,10)                  & (10,10,10)                & (10,10,10)                 \\ 
\hline
B16               & (8,9,10)                  & (8,9,10)                  & (8,9,10)                  & (6,7,8)                    \\
\hline
\end{tabular}
\label{tab:expert_group}
}
\end{adjustwidth}
\end{table*}

\subsubsection*{Defuzzification}
Defuzzification, often referred to as definite value creation, reduces a fuzzy set to a single crisp value. It transforms the result of the fuzzy inference process into a single value that can be utilized as a decision or action, making it a critical step in fuzzy logic systems. There are several techniques to defuzzify a fuzzy integer; however, the centroid method is employed in this stage. Using the geometric mean, this method determines the center of mass of the fuzzy set and produces an associated crisp value.

The following are the fuzzy weights for the barriers:
\begin{equation}
    a_-j = (a_{j}, b_{j}, c_{j})
\end{equation}

where \(a_{j} = \min\{a_{ij}\},\)
\(b_{j} = \sqrt[n]{\prod_{i=1}^{n} b_{ij}},\)
and \(c_{j} = \max\{c_{ij}\}.\)

In this study, the group choices of experts were ascertained using a geometric mean approach.

To obtain a crisp value (\(S_j\)), which is determined by the following formula, the fuzzy weights are defuzzified using the straightforward center of gravity approach:
\begin{equation}
    S_j = \frac{a_j + b_j + c_j}{3}, \quad j \in \{1,2, \ldots, m\}
\end{equation}

\subsubsection*{Barrier selection based on significance}

In the barrier selection process, our objective is to prioritize barriers based on their significance within a given context. This prioritization aids in focusing on resources and efforts to address the most critical obstacles. To operationalize this, we employ a systematic approach involving calculating a threshold value denoted as \(S\), which serves as a criterion for barrier selection. Threshold \(S\) was determined using the formula:

\begin{equation}
S = \frac{\sum_{i=1}^{n} S_i}{n}
\end{equation}

Here, \(i\) ranged from 1 to \(n\), where \(n\) represents the total number of barriers evaluated. The following principles guide the selection of barriers:

\begin{itemize}
    \item If the significance score \(S_i\) of barrier \(i\) is greater than or equal to the threshold \(S\), barrier \(i\) is deemed worthy of further consideration.
    \item Conversely, if \(S_i\) is less than \(S\), barrier \(i\) is excluded from further analysis as it does not meet the significance criterion.
\end{itemize}

The values of \(S\) and the subsequent decision outcomes are summarized in Table \ref{tab5}, where each barrier is evaluated based on its defuzzification weighted value and categorized as either selected or rejected.

\begin{table*}[!ht]
\centering
\caption{\bf Selection of final barriers based on significance.}
\begin{tabular}{|c|c|c|} 
\hline
\textbf{Barriers} & \textbf{Defuzzification Weighted Value} & \textbf{Decision}  \\ 
\hline
B1                & 7.41                                    & Selected           \\ 
\hline
B2                & 8.07                                    & Selected           \\ 
\hline
B3                & 5.99                                    & Rejected           \\ 
\hline
B4                & 8.41                                    & Selected           \\ 
\hline
B5                & 7.62                                    & Selected           \\ 
\hline
B6                & 5.06                                    & Rejected           \\ 
\hline
B7                & 4.84                                    & Rejected           \\ 
\hline
B8                & 3.89                                    & Rejected           \\ 
\hline
B9                & 9.16                                    & Selected           \\ 
\hline
B10               & 7.9                                     & Selected           \\ 
\hline
B11               & 8.06                                    & Selected           \\ 
\hline
B12               & 3.89                                    & Rejected           \\ 
\hline
B13               & 8.22                                    & Selected           \\ 
\hline
B14               & 8.07                                    & Selected           \\ 
\hline
B15               & 9.25                                    & Selected           \\ 
\hline
B16               & 8.15                                    & Selected           \\
\hline
\end{tabular}
\label{tab5}
\end{table*}

\subsection*{FAHP methodology}
The FAHP's pairwise assessments of criteria and alternatives are carried out using linguistic variables represented by triangular numbers. Although FAHP offers a range of calculation methods, we opt for Buckley's FAHP method in this study because of its elegant simplicity and effectiveness in handling the complexity of our research scope. Buckley's FAHP is a modification of Saaty's AHP approach, designed to incorporate uncertainty into a mathematical model while maintaining its representation. The FAHP process unfolds systematically through the following steps:

\textbf{Hierarchy construction:} We begin by meticulously constructing a hierarchy that encapsulates the objectives of the various barriers under consideration, subsequently organizing them into a structured list of options. Each barrier was outlined and defined within the context of our research goal, aiming to discern the most crucial barriers within the selected set. This hierarchical representation provides a clear and organized framework for subsequent evaluations, ensuring that every aspect is considered in the ranking process.

\textbf{Pairwise comparison matrices between barriers:} To rank the barriers in the hierarchy according to their relative importance to one another, fuzzy triangular scales are utilized. As shown in Table \ref{tab6}, the barriers are contrasted using language phrases.

\begin{table*}[!ht]
\begin{adjustwidth}{-2.25in}{0in}
\centering
\caption{\bf Linguistic terms and corresponding triangular fuzzy scale for barrier significance assessment.}
\begin{tabular}{|c|c|c|} 
\hline
\textbf{Saaty Scale} & \textbf{Linguistic Terms}                                                & \textbf{Corresponding Triangular Fuzzy Number}  \\ 
\hline
1                    & \begin{tabular}[c]{@{}c@{}}Equally important\\ ~(Eq. Imp)\end{tabular}   & (1,1,1)                                         \\ 
\hline
3                    & Weakly important (W. Imp)                                                & (2,3,4)                                         \\ 
\hline
5                    & Fairly important (F. Imp)                                                & (4,5,6)                                         \\ 
\hline
7                    & Strongly important (S.Imp)                                               & (6,7,8)                                         \\ 
\hline
9                    & \begin{tabular}[c]{@{}c@{}}Absolutely important\\ ~(A. Imp)\end{tabular} & (9,9,9)                                         \\ 
\hline
2                    & The
  intermittent values between two adjacent scales                    & (1,2,3)                                         \\ 
\hline
4                    & 4.84                                                                     & (3,4,5)                                         \\ 
\hline
6                    & 3.89                                                                     & (5,6,7)                                         \\ 
\hline
8                    & 9.16                                                                     & (7,8,9)                                         \\
\hline
\end{tabular}
\label{tab6}
\end{adjustwidth}
\end{table*}

Leveraging insights gleaned from sets of expert questionnaires, we conducted pairwise comparisons among all the barriers. To consolidate the expert opinions, we calculated the average of the collected data. Subsequently, the data obtained from pairwise comparisons, as presented in Table \ref{tab2}, were aggregated on a triangular scale to form a matrix of contributions. Table \ref{tab7} shows the pairwise comparison matrix of barriers' significance. The elements in each row represent the relative significance of the corresponding barriers concerning one another, as assessed by experts using a triangular scale. Each value reflects the degree of significance, with larger values indicating greater significance. The table assists in prioritizing and understanding the relative importance of barriers in this study.

\begin{table}[!ht]
\begin{adjustwidth}{-2.25in}{0in}
\centering
\caption{\bf Pairwise comparison matrix of barriers' significance.}
\resizebox{\linewidth}{!}{%
\begin{tabular}{|c|c|c|c|c|c|c|c|c|c|c|c|c|c|c|c|} 
\hline
 & \multicolumn{3}{c|}{\textbf{B1}} & \multicolumn{3}{c|}{\textbf{B2}} & \multicolumn{3}{c|}{\textbf{B3}} & \multicolumn{3}{c|}{\textbf{B4}} & \multicolumn{3}{c|}{\textbf{B5}} \\ 
\hline
\textbf{B1} & 1 & 1 & 1 & 0.25 & 0.33 & 0.5 & 0.17 & 0.2 & 0.25 & 2 & 3 & 4 & 0.125 & 0.147 & 0.17 \\ 
\hline
\textbf{B2} & 2 & 3 & 4 & 1 & 1 & 1 & 0.17 & 0.2 & 0.25 & 1 & 1 & 1 & 9 & 9 & 9 \\ 
\hline
\textbf{B3} & 4 & 5 & 6 & 4 & 5 & 6 & 1 & 1 & 1 & 2 & 3 & 4 & 0.17 & 0.2 & 0.25 \\ 
\hline
\textbf{B4} & 0.25 & 0.33 & 0.5 & 1 & 1 & 1 & 0.25 & 0.33 & 0.5 & 1 & 1 & 1 & 0.17 & 0.2 & 0.25 \\ 
\hline
\textbf{B5} & 6 & 7 & 8 & 0.11 & 0.11 & 0.11 & 4 & 5 & 6 & 4 & 5 & 6 & 1 & 1 & 1 \\ 
\hline
\textbf{B6} & 0.17 & 0.2 & 0.25 & 0.25 & 0.33 & 0.5 & 0.25 & 0.33 & 0.5 & 0.25 & 0.33 & 0.5 & 0.25 & 0.33 & 0.5 \\ 
\hline
\textbf{B7} & 1 & 1 & 1 & 4 & 5 & 6 & 2 & 3 & 4 & 2 & 3 & 4 & 4 & 5 & 6 \\ 
\hline
\textbf{B8} & 0.25 & 0.33 & 0.5 & 1 & 1 & 1 & 1 & 1 & 1 & 0.17 & 0.2 & 0.17 & 0.17 & 0.2 & 0.25 \\ 
\hline
\textbf{B9} & 4 & 5 & 6 & 6 & 7 & 8 & 2 & 3 & 4 & 0.11 & 0.11 & 0.11 & 4 & 5 & 6 \\ 
\hline
\textbf{B10} & 9 & 9 & 9 & 2 & 3 & 4 & 4 & 5 & 6 & 2 & 3 & 4 & 0.25 & 0.33 & 0.5 \\ 
\hline
\textbf{B11} & 2 & 3 & 4 & 0.125 & 0.147 & 0.17 & 0.17 & 0.2 & 0.25 & 6 & 7 & 8 & 0.125 & 0.147 & 0.17 \\
\hline
\end{tabular}
}
\label{tab7}
\end{adjustwidth}
\end{table}

\begin{table}[!ht]
\begin{adjustwidth}{-2.25in}{0in}
\centering
\caption*{\bf Continuation of Table \ref{tab7}: Pairwise comparison matrix of barriers' significance.}
\resizebox{\linewidth}{!}{%
\begin{tabular}{|c|c|c|c|c|c|c|c|c|c|c|c|c|c|c|c|c|c|c|} 
\hline
 & \multicolumn{3}{c|}{\textbf{B6}} & \multicolumn{3}{c|}{\textbf{B7}} & \multicolumn{3}{c|}{\textbf{B8}} & \multicolumn{3}{c|}{\textbf{B9}} & \multicolumn{3}{c|}{\textbf{B10}} & \multicolumn{3}{c|}{\textbf{B11}} \\ 
\hline
\textbf{B1} & 4 & 5 & 6 & 1 & 1 & 1 & 2 & 3 & 4 & 0.17 & 0.2 & 0.25 & 0.111 & 0.111 & 0.111 & 0.25 & 0.33 & 0.5 \\ 
\hline
\textbf{B2} & 2 & 3 & 4 & 0.17 & 0.2 & 0.25 & 1 & 1 & 1 & 0.125 & 0.147 & 0.17 & 0.25 & 0.33 & 0.5 & 6 & 7 & 8 \\ 
\hline
\textbf{B3} & 2 & 3 & 4 & 0.25 & 0.33 & 0.5 & 1 & 1 & 1 & 0.25 & 0.33 & 0.5 & 0.17 & 0.2 & 0.25 & 4 & 5 & 6 \\ 
\hline
\textbf{B4} & 2 & 3 & 4 & 0.25 & 0.33 & 0.5 & 4 & 5 & 6 & 9 & 9 & 9 & 0.25 & 0.33 & 0.5 & 0.125 & 0.147 & 0.17 \\ 
\hline
\textbf{B5} & 2 & 3 & 4 & 0.17 & 0.2 & 0.25 & 4 & 5 & 6 & 0.17 & 0.2 & 0.25 & 2 & 3 & 4 & 6 & 7 & 8 \\ 
\hline
\textbf{B6} & 1 & 1 & 1 & 0.25 & 0.33 & 0.5 & 2 & 3 & 4 & 0.25 & 0.33 & 0.5 & 0.125 & 0.142 & 0.17 & 2 & 3 & 4 \\ 
\hline
\textbf{B7} & 2 & 3 & 4 & 1 & 1 & 1 & 1 & 1 & 1 & 0.17 & 0.2 & 0.25 & 1 & 1 & 1 & 4 & 5 & 6 \\ 
\hline
\textbf{B8} & 0.25 & 0.33 & 0.5 & 1 & 1 & 1 & 1 & 1 & 1 & 0.25 & 0.33 & 0.5 & 0.17 & 0.2 & 0.25 & 4 & 5 & 6 \\ 
\hline
\textbf{B9} & 2 & 3 & 4 & 4 & 5 & 6 & 2 & 3 & 4 & 1 & 1 & 1 & 0.125 & 0.147 & 0.17 & 9 & 9 & 9 \\ 
\hline
\textbf{B10} & 6 & 7 & 8 & 1 & 1 & 1 & 4 & 5 & 6 & 6 & 7 & 8 & 1 & 1 & 1 & 2 & 3 & 4 \\ 
\hline
\textbf{B11} & 0.25 & 0.33 & 0.5 & 0.25 & 0.33 & 0.5 & 0.17 & 0.2 & 0.25 & 0.11 & 0.11 & 0.11 & 0.25 & 0.33 & 0.5 & 1 & 1 & 1 \\
\hline
\end{tabular}
}
\end{adjustwidth}
\end{table}

\textbf{Normalized Relative Weights of Barriers:} In this stage, we calculate the normalized relative weights of barriers using the geometric mean of the fuzzy comparison values. An example calculation for ``Criteria 1" is presented in Equation (\ref{eq5}):

\begin{equation}
\tilde{r}_l = \left( \prod_{j=1}^{n} \tilde{d}_{ij} \right)^{\frac{1}{n}}
\label{eq5}
\end{equation}

For ``Criteria 1," the calculation is as follows:

$\tilde{r}_l$ $= [(1 \times 0.25 \times 0.17 \times 2 \times 0.125 \times 4 \times 1 \times 2 \times 0.17 \times 0.111 \times 0.25)^{\frac{1}{11}}; 
(1 \times 0.33 \times 0.2 \times 3 \times 0.147 \times 5 \times 1 \times 3 \times 0.2 \times 0.111 \times 0.33)^{\frac{1}{11}}; 
(1 \times 0.5 \times 0.25 \times 4 \times 0.17 \times 6 \times 1 \times 4 \times 0.25 \times 0.111 \times 0.5)^{\frac{1}{11}}]$

This results in the following normalized relative weights for ``Criteria 1":

\[
\tilde{r}_l = [0.4911, 0.5932, 0.7232]
\]

These values represent the normalized weights for ``Criteria 1" for the given fuzzy comparison values. Table \ref{tab8} presents the geometric means of the fuzzy comparison values (ri) for various criteria labeled B1 to B11. These values indicate the normalized relative weights for each criterion. Additionally, the table includes a ``Total" row summarizing the cumulative values for all criteria and two additional rows denoted as ``P(-1)" and ``INCR," which represent specific parameters or calculations related to the analysis. These values played a crucial role in assessing the relative importance of the criteria within the study context.

\begin{table}
\centering
\caption{\bf Geometric means of fuzzy comparison values ($r_i$) for criteria B1 to B11.}
\begin{tabular}{|c|c|c|c|} 
\hline
\multicolumn{4}{|c|}{\textbf{~~ Geometric mean of fuzzy}} \\ 
\hline
~ & \multicolumn{3}{c|}{\textbf{$r_i$}} \\ 
\hline
\textbf{B1} & 0.4911269 & 0.593166 & 0.723203 \\ 
\hline
\textbf{B2} & 0.8619317 & 1.008719 & 1.179025 \\ 
\hline
\textbf{B3} & 0.9322815 & 1.155445 & 1.437111 \\ 
\hline
\textbf{B4} & 0.6278548 & 0.757425 & 0.950749 \\ 
\hline
\textbf{B5} & 1.3594037 & 1.647404 & 1.946593 \\ 
\hline
\textbf{B6} & 0.3752132 & 0.482518 & 0.661569 \\ 
\hline
\textbf{B7} & 1.5008527 & 1.808046 & 2.097435 \\ 
\hline
\textbf{B8} & 0.4793547 & 0.551569 & 0.644492 \\ 
\hline
\textbf{B9} & 1.7160993 & 2.096448 & 2.444355 \\ 
\hline
\textbf{B10} & 2.3177318 & 2.843423 & 3.382683 \\ 
\hline
\textbf{B11} & 0.3488563 & 0.420009 & 0.52264 \\ 
\hline
\textbf{Total} & 11.010707 & 13.36417 & 15.98986 \\ 
\hline
\textbf{P(-1)} & 0.0908207 & 0.074827 & 0.06254 \\ 
\hline
\textbf{INCR} & 0.0625397 & 0.111 & 0.090821 \\
\hline
\end{tabular}
\label{tab8}
\end{table}

In Table \ref{tab9}, the relative fuzzy weights for each barrier are calculated using a method that combines the geometric means of fuzzy values with the sum of the geometric means of reverse fuzzy values in ascending order, as follows: $\tilde w_l = [ (0.4911269 \times 0.0625397), (0.593166 \times 0.111), (0.723203 \times 0.090821)] = [0.0307, 0.0658, 0.0657]$. The resulting values represent the relative importance of each barrier.

\begin{table}
\centering
\caption{\bf Relative fuzzy weight of each barrier.}
\begin{tabular}{|c|c|c|c|} 
\hline
\multicolumn{4}{|c|}{\textbf{Fuzzy weight}} \\ 
\hline
\textbf{~} & \multicolumn{3}{c|}{\textbf{$W_i$}} \\ 
\hline
\textbf{B1} & 0.0307 & 0.0658 & 0.0657 \\ 
\hline
\textbf{B2} & 0.0539 & 0.112 & 0.1071 \\ 
\hline
\textbf{B3} & 0.0583 & 0.1283 & 0.1305 \\ 
\hline
\textbf{B4} & 0.0393 & 0.0841 & 0.0863 \\ 
\hline
\textbf{B5} & 0.085 & 0.1829 & 0.1768 \\ 
\hline
\textbf{B6} & 0.0235 & 0.0536 & 0.0601 \\ 
\hline
\textbf{B7} & 0.0939 & 0.2007 & 0.1905 \\ 
\hline
\textbf{B8} & 0.03 & 0.0612 & 0.0585 \\ 
\hline
\textbf{B9} & 0.1073 & 0.2327 & 0.222 \\ 
\hline
\textbf{B10} & 0.145 & 0.3156 & 0.3072 \\ 
\hline
\textbf{B11} & 0.0218 & 0.0466 & 0.0475 \\
\hline
\end{tabular}
\label{tab9}
\end{table}

The fuzzy numbers associated with each barrier were aggregated to calculate the relative nonfuzzy weights for each barrier. The normalized weights for each barrier were determined by dividing the relative fuzzy weight for each barrier by the sum of the fuzzy weights for all barriers. Table \ref{tab10} presents the average and normalized weights of these barriers.

\begin{table}
\centering
\caption{\bf Averaged ($M_i$) and normalized ($N_i$) relative weights of criteria.}
\begin{tabular}{|c|c|c|} 
\hline
 & \textbf{ $M_i$ } & \textbf{ $N_i$ } \\ 
\hline
\textbf{ B1 } & 0.054 & 0.04476 \\ 
\hline
\textbf{ B2 } & 0.091 & 0.07531 \\ 
\hline
\textbf{ B3 } & 0.106 & 0.08749 \\ 
\hline
\textbf{ B4 } & 0.07 & 0.05786 \\ 
\hline
\textbf{ B5 } & 0.148 & 0.12269 \\ 
\hline
\textbf{ B6 } & 0.046 & 0.03783 \\ 
\hline
\textbf{ B7 } & 0.162 & 0.13383 \\ 
\hline
\textbf{ B8 } & 0.05 & 0.04132 \\ 
\hline
\textbf{ B9 } & 0.187 & 0.15507 \\ 
\hline
\textbf{ B10 } & 0.256 & 0.21185 \\ 
\hline
\textbf{ B11 } & 0.039 & 0.03198 \\ 
\hline
\textbf{ Total } & 1.208 & 1 \\
\hline
\end{tabular}
\label{tab10}
\end{table}

\section*{Results and discussions}
\label{results}
Table \ref{tab11} presents the priority weights for barriers to IoT adoption in the Bangladeshi manufacturing industry, calculated using FAHP.

\begin{table*}[!ht]
\begin{adjustwidth}{-2.25in}{0in}
\centering
\caption{\bf Relative weight and ranking of IoT adoption barriers in the Bangladeshi industrial sector.}
\begin{tabular}{|c|l|c|c|}
\hline
\textbf{Barrier No} & \textbf{Barriers to IoT Adoption} & \textbf{Weights} & \textbf{Ranking} \\
\hline
B1 & Absence of training initiatives & 0.04476 & 8 \\
B2 & Lack of knowledge of ROI & 0.07531 & 6 \\
B3 & Unawareness of IoT advantages & 0.08749 & 5 \\
B4 & Employees' resistance to new tech & 0.05786 & 7 \\
B5 & Security and privacy concerns & 0.12269 & 4 \\
B6 & Unskilled or inexperienced Manpower & 0.03783 & 10 \\
B7 & Risks with new business model & 0.13383 & 3 \\
B8 & Inadequate technology & 0.04132 & 9 \\
B9 & High initial implementation costs & 0.15507 & 2 \\
B10 & Lack of top management commitment & 0.21185 & 1 \\
B11 & Absence of Standardization & 0.03198 & 11 \\
\hline
\end{tabular}
\label{tab11}
\end{adjustwidth}
\end{table*}

The most critical barrier identified through our research is the ``lack of top management's commitment to implementing new technology" (B10), with a weight of 0.21185. Other significant barriers include the ``high initial implementation investment costs" (B9) at 0.15507 and ``Risks associated with switching to a new business model" (B7) at 0.13383. These three factors require immediate attention for more effective IoT adoption in the Bangladeshi manufacturing industry. Top management must understand the adoption of new technology despite the significant initial investment costs. Furthermore, adopting new technology necessitates changes in business strategies and a well-planned action plan. Organizations may be reluctant to embrace such changes because of their high risk and potential financial losses.

Ranking from fourth to seventh is considered a barrier of intermediate importance. ``Security and privacy concerns" (B5) rank fourth at 0.12269, ``Unawareness of the advantages of IoT" (B3) is fifth at 0.08749, ``Lack of knowledge of return on investment (ROI)" (B2) is sixth at 0.07531, and ``Employees' resistance to embracing new technologies" (B4) is seventh at 0.05786. However, these barriers are significant but relatively less substantial than the top-tier obstacles. Privacy and security issues are well-documented in the existing literature. A comprehensive understanding of IoT benefits and the conventional payback model is essential to maximize the returns on IoT investment. Organizations should also help employees understand how their roles will change by adopting new technologies to reduce resistance.

By contrast, the remaining barriers have lower rankings and less impact on IoT adoption decisions in the manufacturing sector. Ranging from eighth to eleventh in relative weights, the least essential barriers are: ``Absence of training initiatives" (B1) with a weight of 0.04476, ``Inadequate technology" (B8) at 0.04132, ``Unskilled or inexperienced Manpower" (B6) at 0.03783, and ``Absence of Standardization" (B11) at 0.03198. There's a clear need to develop the technology necessary for IoT adoption in the Bangladeshi manufacturing sector. Government agencies and policymakers should focus on creating essential technology infrastructure. Specialized training programs are necessary to equip personnel with the skills required for IoT adoption.

\section*{Managerial implications}\label{sec:managerial}
This study employs a rigorous methodology integrating literature analysis, expert opinions, and analytical hierarchy processing to offer manufacturing leaders and policymakers in Bangladesh critical insights into the obstacles hindering the adoption of IoT. By pinpointing concerns about minimal C-suite backing, high costs, and aversion to operational shifts using advanced fuzzy prioritization, executives can align strategic priorities to tackle the most pressing challenges first.

Specifically, factory heads should channel resources toward fostering a culture that is open to top-down technological integration. Building financial justification for IoT by projecting a return on investment under various implementation scenarios can secure an initial pilot investment sign-off. Driving change management programs that contain organizational resistance is imperative. Additionally, forging public-private consortiums could drive the development of technical capabilities and infrastructure. Architectural standardization and policy incentives also play key roles over time.

Armed with clarity on specific bottlenecks curbing the assimilation of connected technologies in their unique context, Bangladeshi manufacturing leaders have an actionable roadmap. They can direct interventions toward tackling the identified impediments using inputs from methodical academic inquiry and data-based analytics. The insights distilled in this study should thereby expedite the adoption of the IoT to unlock efficiency gains, visibility, and competitive edge for the sector. This aligns well with Vision 2041's digital transformation goals and private sector-led economic growth.

\section*{Real-world applications and practical implications}\label{sec:real-world}
Our research on the challenges hindering the adoption of Internet of Things (IoT) technologies in the manufacturing sector has significant implications for real-world applications and practical decision-making. By systematically identifying and evaluating these challenges, our study offers valuable insights that can inform strategic decision-making and facilitate the successful implementation of IoT initiatives in manufacturing firms.

One compelling example of IoT implementation in manufacturing is shown by Siemens AG, a multinational conglomerate operating in various industries, including manufacturing. Siemens implemented IoT-enabled predictive maintenance solutions in its factories by leveraging real-time data from sensors embedded in machinery to anticipate equipment failures before they occur. This proactive maintenance approach reduces downtime and optimizes production schedules, resulting in significant cost savings and operational efficiency improvements.

Additionally, we consider the case study of General Electric (GE), a renowned manufacturer of industrial equipment and machinery. GE utilizes IoT technologies to enhance its supply chain visibility and optimize inventory management processes. By deploying RFID tags and IoT-enabled tracking systems, GE gained real-time insights into inventory levels and shipment status, enabling more accurate demand forecasting and the timely replenishment of materials. As a result, the company streamlined its supply chain operations, minimized stockouts, and improved its overall operational performance.

The practical implications of our research are manifold and hold significant value for academia and industry practitioners. First, it provides a comprehensive understanding of the challenges hindering the adoption of IoT technologies in the manufacturing sector. By systematically identifying and evaluating these barriers, we offer insights into the complex terrain of IoT adoption, enabling stakeholders to anticipate and proactively address potential obstacles. Our proposed methodology, which combines the Delphi method with FAHP, offers a practical framework for analyzing and prioritizing IoT adoption barriers. This methodology provides a structured approach for decision-makers to systematically assess the challenges specific to their organizational context and allocate resources accordingly. For instance, a case study conducted by IBM illustrated the application of FAHP in prioritizing IoT adoption barriers in a manufacturing setting. By leveraging the FAHP, IBM was able to identify key obstacles and develop targeted strategies to address them, ultimately facilitating the successful implementation of IoT initiatives. Moreover, our research highlights the importance of strategic planning and stakeholder engagement in facilitating successful IoT adoption. By emphasizing the need for top management commitment and collaboration across departments, our findings underscore the critical role of leadership and organizational culture in driving transformative change.

Our study offers practical recommendations for overcoming IoT adoption barriers tailored to manufacturing firms' unique challenges. These recommendations encompass strategies for addressing issues such as high initial implementation costs, technological complexity, and resistance to change, providing actionable insights for industry practitioners seeking to navigate the IoT adoption journey effectively.

The practical implications of our research extend beyond academia, serving as a pragmatic guide for industrial managers and decision-makers. Armed with the knowledge gleaned from our study, stakeholders can develop tailored strategies, set informed priorities, and embark on a transformational journey towards harnessing the vast potential of IoT technologies in the manufacturing sector. By facilitating the adoption of IoT technologies, our research contributes to enhancing operational efficiency, driving innovation, and maintaining competitiveness in an increasingly digitalized and interconnected world.

\section*{Conclusions and future works}
\label{conclusion}
In this study, we comprehensively examined the barriers that manufacturing firms face in embracing IoT technologies. By leveraging the combined Delphi and FAHP methodology, we identified these challenges and ranked them in terms of their relative significance,  illuminating the landscape of IoT adoption within the context of the Bangladeshi manufacturing industry. Our study yielded several key insights. First, we identified and ranked 11 prominent barriers to IoT adoption in the Bangladeshi manufacturing industry. These findings serve as a valuable resource for industry practitioners and decision-makers, providing them with a clear understanding of the challenges they must address. Our study highlights that ``Lack of top management commitment to implementing new technology" (B10), ``High initial implementation investment costs" (B9), and ``Risks associated with switching to a new business model" (B7) are the top three hurdles that require immediate attention. Furthermore, our research contributes to the growing body of knowledge on IoT adoption in the manufacturing sector, shedding light on nuanced challenges specific to the Bangladeshi context. This underscores the need for tailored strategies and initiatives to facilitate a smooth transition to IoT technologies.

It is important to acknowledge the limitations of this study. It primarily focuses on barriers, potentially overlooking opportunities and enablers. The findings are specific to Bangladeshi manufacturing, and their generalizability to other regions and sectors may require further investigation. Additionally, the FAHP approach, while effective, relies on expert opinions and predefined criteria, thereby introducing subjectivity.

Future research should develop practical strategies to address these identified barriers, involving in-depth case studies and real-world implementation. Expanding the scope to different regions and sectors would offer a more comprehensive view while investigating the enablers and success factors, which could provide valuable insights. Ongoing research is crucial for understanding the emerging challenges and opportunities in the ever-evolving IoT landscape.

Our study lays the foundation for understanding and addressing IoT adoption challenges in manufacturing firms. Our understanding and strategies must evolve as technology advances to ensure successful IoT integration in the manufacturing sector.

\nolinenumbers

%
%
%

\end{justify}

\end{document}